\documentclass[11pt]{article}
\usepackage[a4paper]{geometry}
\usepackage{amsfonts, amsmath, amssymb, amsthm, graphicx, caption, authblk, multirow, makecell, framed, float, xcolor, enumitem, tikz, hyperref}
\usepackage{rotating}
\definecolor{blk}{RGB}{63,63,63}
\theoremstyle{definition}
\newtheorem{theorem}{Theorem}

\theoremstyle{remark}

\newcommand*{\mybox}[1]{%
  \framebox{\raisebox{0cm}[0.5\baselineskip][0.05\baselineskip]{%
    \hbox to 0.10cm {\hss#1\hss}}}\hspace{0.05cm}}
\newcommand{\mystack}[1]{\begin{array}{|r|||}\hline#1\\\hline\end{array}}

\begin{document}
\title{Nondango is NP-Complete}
\author[1]{Suthee Ruangwises\thanks{\texttt{ruangwises@gmail.com}}}
\affil[1]{Department of Informatics, The University of Electro-Communications, Tokyo, Japan}
\date{}
\maketitle

\begin{abstract}
Nondango is a pencil puzzle consisting of a rectangular grid partitioned into regions, with some cells containing a white circle. The player has to color some circles black such that every region contains exactly one black circle, and there are no three consecutive circles (horizontally, vertically, or diagonally) having the same color. In this paper, we prove that deciding solvability of a given Nondango puzzle is NP-complete.

\end{abstract}

\section{Introduction}
\begin{figure}
\centering
\begin{tikzpicture}
\draw[step=0.8cm,color={rgb:black,1;white,4}] (0,0) grid (4.8,4.8);

\draw[line width=0.6mm] (0,0) -- (0,4.8);
\draw[line width=0.6mm] (4.8,0) -- (4.8,4.8);
\draw[line width=0.6mm] (0,0) -- (4.8,0);
\draw[line width=0.6mm] (0,4.8) -- (4.8,4.8);

\draw[line width=0.6mm] (0.8,0) -- (0.8,2.4);
\draw[line width=0.6mm] (0.8,3.2) -- (0.8,4);
\draw[line width=0.6mm] (1.6,0.8) -- (1.6,2.4);
\draw[line width=0.6mm] (1.6,3.2) -- (1.6,4.8);
\draw[line width=0.6mm] (2.4,0) -- (2.4,2.4);
\draw[line width=0.6mm] (2.4,3.2) -- (2.4,4.8);
\draw[line width=0.6mm] (3.2,0.8) -- (3.2,4);
\draw[line width=0.6mm] (4,1.6) -- (4,4);

\draw[line width=0.6mm] (0.8,0.8) -- (1.6,0.8);
\draw[line width=0.6mm] (3.2,0.8) -- (4.8,0.8);
\draw[line width=0.6mm] (0,1.6) -- (0.8,1.6);
\draw[line width=0.6mm] (2.4,1.6) -- (3.2,1.6);
\draw[line width=0.6mm] (4,1.6) -- (4.8,1.6);
\draw[line width=0.6mm] (0.8,2.4) -- (2.4,2.4);
\draw[line width=0.6mm] (3.2,2.4) -- (4,2.4);
\draw[line width=0.6mm] (0.8,3.2) -- (2.4,3.2);
\draw[line width=0.6mm] (0,4) -- (0.8,4);
\draw[line width=0.6mm] (2.4,4) -- (4,4);

\node[draw,circle,minimum size=0.4cm] at (2,0.4) {};
\node[draw,circle,minimum size=0.4cm] at (2.8,0.4) {};
\node[draw,circle,minimum size=0.4cm] at (3.6,0.4) {};
\node[draw,circle,minimum size=0.4cm] at (0.4,1.2) {};
\node[draw,circle,minimum size=0.4cm] at (1.2,1.2) {};
\node[draw,circle,minimum size=0.4cm] at (2,1.2) {};
\node[draw,circle,minimum size=0.4cm] at (3.6,1.2) {};
\node[draw,circle,minimum size=0.4cm] at (0.4,2) {};
\node[draw,circle,minimum size=0.4cm] at (2.8,2) {};
\node[draw,circle,minimum size=0.4cm] at (3.6,2) {};
\node[draw,circle,minimum size=0.4cm] at (4.4,2) {};
\node[draw,circle,minimum size=0.4cm] at (0.4,2.8) {};
\node[draw,circle,minimum size=0.4cm] at (3.6,2.8) {};
\node[draw,circle,minimum size=0.4cm] at (0.4,3.6) {};
\node[draw,circle,minimum size=0.4cm] at (1.2,3.6) {};
\node[draw,circle,minimum size=0.4cm] at (2,3.6) {};
\node[draw,circle,minimum size=0.4cm] at (2.8,3.6) {};
\node[draw,circle,minimum size=0.4cm] at (3.6,3.6) {};
\node[draw,circle,minimum size=0.4cm] at (4.4,3.6) {};
\end{tikzpicture}
\hspace{1.5cm}
\begin{tikzpicture}
\draw[step=0.8cm,color={rgb:black,1;white,4}] (0,0) grid (4.8,4.8);

\draw[line width=0.6mm] (0,0) -- (0,4.8);
\draw[line width=0.6mm] (4.8,0) -- (4.8,4.8);
\draw[line width=0.6mm] (0,0) -- (4.8,0);
\draw[line width=0.6mm] (0,4.8) -- (4.8,4.8);

\draw[line width=0.6mm] (0.8,0) -- (0.8,2.4);
\draw[line width=0.6mm] (0.8,3.2) -- (0.8,4);
\draw[line width=0.6mm] (1.6,0.8) -- (1.6,2.4);
\draw[line width=0.6mm] (1.6,3.2) -- (1.6,4.8);
\draw[line width=0.6mm] (2.4,0) -- (2.4,2.4);
\draw[line width=0.6mm] (2.4,3.2) -- (2.4,4.8);
\draw[line width=0.6mm] (3.2,0.8) -- (3.2,4);
\draw[line width=0.6mm] (4,1.6) -- (4,4);

\draw[line width=0.6mm] (0.8,0.8) -- (1.6,0.8);
\draw[line width=0.6mm] (3.2,0.8) -- (4.8,0.8);
\draw[line width=0.6mm] (0,1.6) -- (0.8,1.6);
\draw[line width=0.6mm] (2.4,1.6) -- (3.2,1.6);
\draw[line width=0.6mm] (4,1.6) -- (4.8,1.6);
\draw[line width=0.6mm] (0.8,2.4) -- (2.4,2.4);
\draw[line width=0.6mm] (3.2,2.4) -- (4,2.4);
\draw[line width=0.6mm] (0.8,3.2) -- (2.4,3.2);
\draw[line width=0.6mm] (0,4) -- (0.8,4);
\draw[line width=0.6mm] (2.4,4) -- (4,4);

\node[draw,circle,minimum size=0.4cm, fill=blk] at (2,0.4) {};
\node[draw,circle,minimum size=0.4cm, fill=blk] at (2.8,0.4) {};
\node[draw,circle,minimum size=0.4cm] at (3.6,0.4) {};
\node[draw,circle,minimum size=0.4cm, fill=blk] at (0.4,1.2) {};
\node[draw,circle,minimum size=0.4cm, fill=blk] at (1.2,1.2) {};
\node[draw,circle,minimum size=0.4cm] at (2,1.2) {};
\node[draw,circle,minimum size=0.4cm] at (3.6,1.2) {};
\node[draw,circle,minimum size=0.4cm] at (0.4,2) {};
\node[draw,circle,minimum size=0.4cm] at (2.8,2) {};
\node[draw,circle,minimum size=0.4cm, fill=blk] at (3.6,2) {};
\node[draw,circle,minimum size=0.4cm] at (4.4,2) {};
\node[draw,circle,minimum size=0.4cm, fill=blk] at (0.4,2.8) {};
\node[draw,circle,minimum size=0.4cm, fill=blk] at (3.6,2.8) {};
\node[draw,circle,minimum size=0.4cm] at (0.4,3.6) {};
\node[draw,circle,minimum size=0.4cm, fill=blk] at (1.2,3.6) {};
\node[draw,circle,minimum size=0.4cm, fill=blk] at (2,3.6) {};
\node[draw,circle,minimum size=0.4cm] at (2.8,3.6) {};
\node[draw,circle,minimum size=0.4cm] at (3.6,3.6) {};
\node[draw,circle,minimum size=0.4cm, fill=blk] at (4.4,3.6) {};
\end{tikzpicture}
\caption{An example of a $6 \times 6$ Nondango puzzle (left) and its solution (right)}
\label{fig0}
\end{figure}
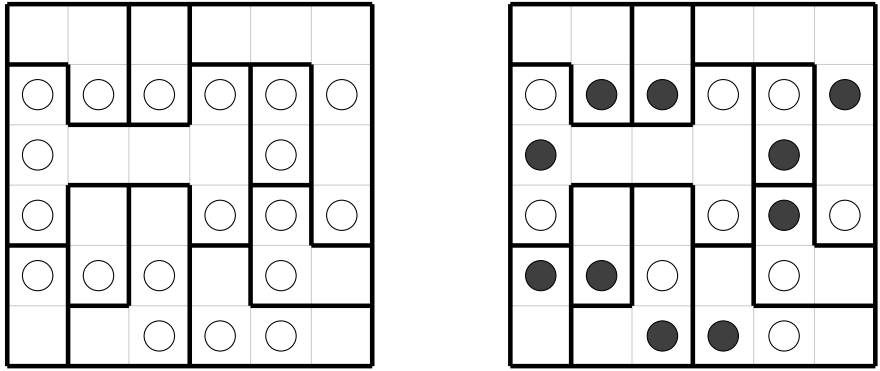

\textit{Nondango} is a pencil puzzle published by Nikoli. The puzzle consists of a rectangular grid partitioned into polyominoes called \textit{regions}, with some cells containing a white circle. The player has to color some circles black to satisfy the following constraints \cite{janko}.
\begin{enumerate}
	\item Every region contains exactly one black circle.
	\item There are no three consecutive circles (horizontally, vertically, or diagonally) having the same color (see Figure \ref{fig0}).
\end{enumerate}

In this paper, we show that it is NP-complete to decide whether a given Nondango puzzle has a solution.

\begin{theorem}
Deciding solvability of a given Nondango instance is NP-complete.
\end{theorem}

As the problem clearly belongs to NP, the nontrivial part is to prove the NP-hardness. We do so by constructing a reduction from the 1-in-3-SAT+ problem (deciding whether there is a Boolean assignment such that every clause has exactly one literal that evaluates to true, in a setting where each clause contains exactly three positive literals), which is known to be NP-complete \cite{3sat}.

\subsection{Related Work}
Many pencil puzzles have been proved to be NP-complete, including Dosun-Fuwari \cite{dosun}, Fillmat \cite{fillmat}, Five Cells \cite{fivecells}, Goishi Hiroi \cite{goishi}, Hashiwokakero \cite{bridges}, Herugolf \cite{makaro}, Heyawake \cite{heyawake}, Juosan \cite{juosan}, Kakuro \cite{sudoku}, Kurodoko \cite{kurodoko}, Kurotto \cite{juosan}, LITS \cite{lits}, Makaro \cite{makaro}, Moon-on-Sun \cite{moon}, Nagareru \cite{moon}, Nonogram \cite{nonogram}, Norinori \cite{lits}, Numberlink \cite{numberlink}, Nurikabe \cite{nurikabe}, Nurimeizu \cite{moon}, Nurimisaki \cite{nurimisaki}, Pencils \cite{pencils}, Ripple Effect \cite{ripple}, Roma \cite{roma}, Sashigane \cite{nurimisaki}, Shakashaka \cite{shakashaka}, Shikaku \cite{ripple}, Slitherlink \cite{sudoku}, Sto-Stone \cite{stone}, Sudoku \cite{sudoku}, Suguru \cite{suguru}, Sumplete \cite{sumplete}, Tatamibari \cite{tatamibari}, Tilepaint \cite{sudoku}, Toichika \cite{toichika}, Usowan \cite{usowan}, Yin-Yang \cite{yinyang}, and Yosenabe \cite{yosenabe}.

\section{Idea of the Proof}
\begin{figure}
\centering
\begin{tikzpicture}
\draw[step=0.8cm,color={rgb:black,1;white,4}] (0,0) grid (12,9.6);

\draw[line width=0.6mm] (0,0.8) -- (12,0.8);
\draw[line width=0.6mm] (0,1.6) -- (12,1.6);
\draw[line width=0.6mm] (0,3.2) -- (12,3.2);
\draw[line width=0.6mm] (0,4) -- (12,4);
\draw[line width=0.6mm] (0,5.6) -- (12,5.6);
\draw[line width=0.6mm] (0,6.4) -- (12,6.4);
\draw[line width=0.6mm] (0,8) -- (12,8);
\draw[line width=0.6mm] (0,8.8) -- (12,8.8);

\draw[line width=0.6mm] (0,0.8) -- (0,1.6);
\draw[line width=0.6mm] (0,3.2) -- (0,4);
\draw[line width=0.6mm] (0,5.6) -- (0,6.4);
\draw[line width=0.6mm] (0,8) -- (0,8.8);
\draw[line width=0.6mm] (12,0.8) -- (12,1.6);
\draw[line width=0.6mm] (12,3.2) -- (12,4);
\draw[line width=0.6mm] (12,5.6) -- (12,6.4);
\draw[line width=0.6mm] (12,8) -- (12,8.8);

\node at (-0.4,1.2) {$C_4$};
\node at (-0.4,2.5) {$\vdots$};
\node at (-0.4,3.6) {$C_3$};
\node at (-0.4,4.9) {$\vdots$};
\node at (-0.4,6) {$C_2$};
\node at (-0.4,7.3) {$\vdots$};
\node at (-0.4,8.4) {$C_1$};

\node at (1.2,9.8) {$x_1$};
\node at (2.4,9.8) {$\dots$};
\node at (3.6,9.8) {$x_2$};
\node at (4.8,9.8) {$\dots$};
\node at (6,9.8) {$x_3$};
\node at (7.2,9.8) {$\dots$};
\node at (8.4,9.8) {$x_4$};
\node at (9.6,9.8) {$\dots$};
\node at (10.8,9.8) {$x_5$};

\node at (6,2.4) {\LARGE some gadgets};
\node at (6,4.8) {\LARGE some gadgets};
\node at (6,7.2) {\LARGE some gadgets};

\node[draw,circle,minimum size=0.4cm] at (1.2,1.2) {};
\node[draw,circle,minimum size=0.4cm] at (3.6,1.2) {};
\node[draw,circle,minimum size=0.4cm] at (10.8,1.2) {};
\node[draw,circle,minimum size=0.4cm] at (6,3.6) {};
\node[draw,circle,minimum size=0.4cm] at (8.4,3.6) {};
\node[draw,circle,minimum size=0.4cm] at (10.8,3.6) {};
\node[draw,circle,minimum size=0.4cm] at (3.6,6) {};
\node[draw,circle,minimum size=0.4cm] at (6,6) {};
\node[draw,circle,minimum size=0.4cm] at (10.8,6) {};
\node[draw,circle,minimum size=0.4cm] at (1.2,8.4) {};
\node[draw,circle,minimum size=0.4cm] at (3.6,8.4) {};
\node[draw,circle,minimum size=0.4cm] at (8.4,8.4) {};
\end{tikzpicture}
\caption{Basic structure of a Nondango instance transformed from a formula consisting of clauses $C_1=x_1 \vee x_2 \vee x_4$, $C_2=x_2 \vee x_3 \vee x_5$, $C_3=x_3 \vee x_4 \vee x_5$, and $C_4=x_1 \vee x_2 \vee x_5$}
\label{fig1}
\end{figure}
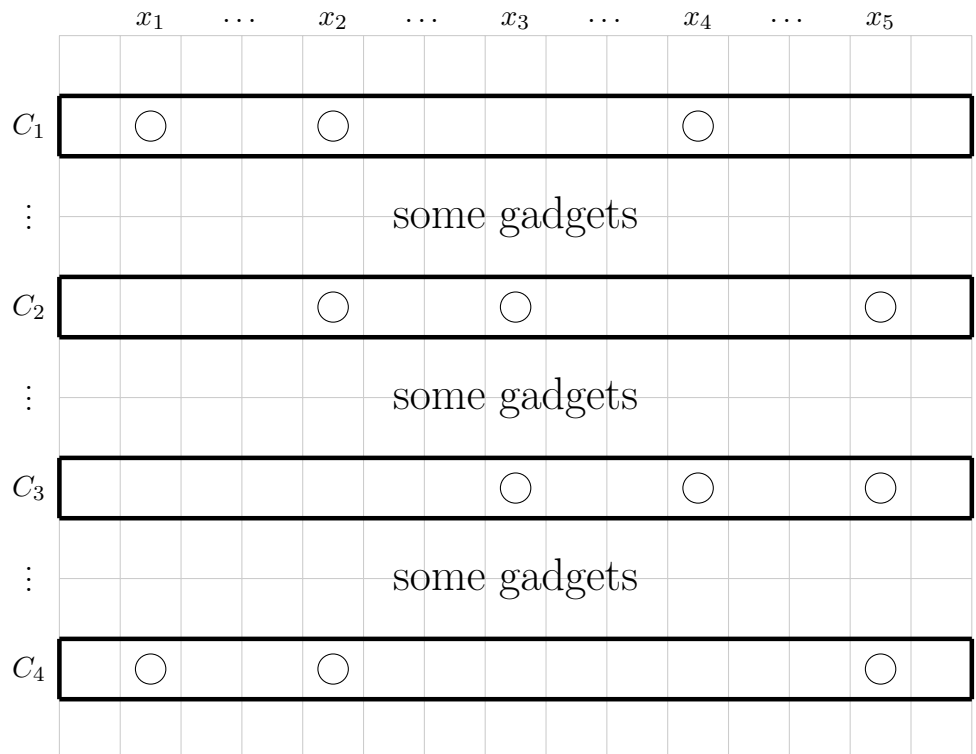

Given a 1-in-3-SAT+ formula, we will transform it into a Nondango puzzle. In the puzzle grid, each variable and each clause is represented by a column (called a \textit{variable column}) and a row (called a \textit{clause row}), respectively. In each clause row, a rectangular region of height 1 consists of the whole row. Inside the region, we place three circles (called \textit{variable circles}) at columns corresponding to the three variables appearing in that clause (see Figure \ref{fig1}).

We interpret a black (resp. white) circle in a Nondango solution as a true (resp. false) literal. The constraint that exactly one literal in each clause is true is equivalent to that exactly one circle in that region is black. However, a more challenging task is to force every circle in each variable column to have the same color (which is equivalent to that each variable must have the same truth value in every clause it appears). We will show how to construct gadgets to enforce this constraint in the next section.

\section{Reduction} \label{gadget}
We use the following gadgets to construct components of the Nondango puzzle with certain properties.

\subsection{Enforcing a Black Circle}
Creating a circle that must be black in the solution is trivial; in a region with exactly one circle, that circle must be black in the solution. We call such circle a \textit{forced black circle}.

\subsection{Enforcing a White Circle}
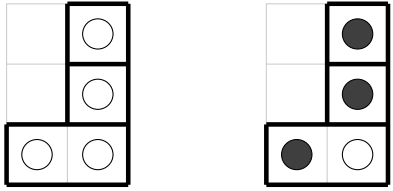
\begin{figure}
\centering
\begin{tikzpicture}
\draw[step=0.8cm,color={rgb:black,1;white,4}] (0.8,2.4) grid (2.4,4.8);

\draw[line width=0.6mm] (0.8,2.4) -- (2.4,2.4);
\draw[line width=0.6mm] (0.8,3.2) -- (2.4,3.2);
\draw[line width=0.6mm] (1.6,4) -- (2.4,4);
\draw[line width=0.6mm] (1.6,4.8) -- (2.4,4.8);

\draw[line width=0.6mm] (0.8,2.4) -- (0.8,3.2);
\draw[line width=0.6mm] (1.6,3.2) -- (1.6,4.8);
\draw[line width=0.6mm] (2.4,2.4) -- (2.4,4.8);

\node[draw,circle,minimum size=0.4cm] at (1.2,2.8) {};
\node[draw,circle,minimum size=0.4cm] at (2,2.8) {};
\node[draw,circle,minimum size=0.4cm] at (2,3.6) {};
\node[draw,circle,minimum size=0.4cm] at (2,4.4) {};
\end{tikzpicture}
\hspace{1.5cm}
\begin{tikzpicture}
\draw[step=0.8cm,color={rgb:black,1;white,4}] (0.8,2.4) grid (2.4,4.8);

\draw[line width=0.6mm] (0.8,2.4) -- (2.4,2.4);
\draw[line width=0.6mm] (0.8,3.2) -- (2.4,3.2);
\draw[line width=0.6mm] (1.6,4) -- (2.4,4);
\draw[line width=0.6mm] (1.6,4.8) -- (2.4,4.8);

\draw[line width=0.6mm] (0.8,2.4) -- (0.8,3.2);
\draw[line width=0.6mm] (1.6,3.2) -- (1.6,4.8);
\draw[line width=0.6mm] (2.4,2.4) -- (2.4,4.8);

\node[draw,circle,minimum size=0.4cm, fill=blk] at (1.2,2.8) {};
\node[draw,circle,minimum size=0.4cm] at (2,2.8) {};
\node[draw,circle,minimum size=0.4cm, fill=blk] at (2,3.6) {};
\node[draw,circle,minimum size=0.4cm, fill=blk] at (2,4.4) {};
\end{tikzpicture}
\caption{A gadget for creating a forced white circle (left) and its only solution (right)}
\label{fig2}
\end{figure}

We can create a circle that must be white in the solution by using a gadget represented in Figure \ref{fig2}. As the bottom-right circle is placed next to two vertically consecutive forced black circles, it must be white in the solution (otherwise there will be three consecutive black circles in the solution). Analogously, we call such circle a \textit{forced white circle}.

\subsection{Enforcing Two Circles with Different Colors} \label{diff}
\begin{figure}
\centering
\begin{tikzpicture}
\draw[step=0.8cm,color={rgb:black,1;white,4}] (0.8,0) grid (4.8,4.8);

\draw[line width=0.6mm] (4,0) -- (4.8,0);
\draw[line width=0.6mm] (0,0.8) -- (5.6,0.8);
\draw[line width=0.6mm] (0,1.6) -- (5.6,1.6);
\draw[line width=0.6mm] (0.8,2.4) -- (2.4,2.4);
\draw[line width=0.6mm] (0.8,3.2) -- (2.4,3.2);
\draw[line width=0.6mm] (1.6,4) -- (2.4,4);
\draw[line width=0.6mm] (1.6,4.8) -- (2.4,4.8);

\draw[line width=0.6mm] (0.8,2.4) -- (0.8,3.2);
\draw[line width=0.6mm] (1.6,3.2) -- (1.6,4.8);
\draw[line width=0.6mm] (2.4,1.6) -- (2.4,5.6);
\draw[line width=0.6mm] (3.2,1.6) -- (3.2,5.6);
\draw[line width=0.6mm] (4,0) -- (4,0.8);
\draw[line width=0.6mm] (4.8,0) -- (4.8,0.8);

\node at (0,1.2) {$\dots$};
\node at (5.6,1.2) {$\dots$};
\node at (2.8,5.7) {$\vdots$};

\node[draw,circle,minimum size=0.4cm] at (4.4,0.4) {};
\node[draw,circle,minimum size=0.4cm] at (3.6,1.2) {};
\node[draw,circle,minimum size=0.4cm] at (2.8,2) {};
\node[draw,circle,minimum size=0.4cm] at (1.2,2.8) {};
\node[draw,circle,minimum size=0.4cm] at (2,2.8) {};
\node[draw,circle,minimum size=0.4cm] at (2,3.6) {};
\node[draw,circle,minimum size=0.4cm] at (2,4.4) {};
\end{tikzpicture}
\hspace{1cm}
\begin{tikzpicture}
\draw[step=0.8cm,color={rgb:black,1;white,4}] (0.8,0) grid (4.8,4.8);

\draw[line width=0.6mm] (4,0) -- (4.8,0);
\draw[line width=0.6mm] (0,0.8) -- (5.6,0.8);
\draw[line width=0.6mm] (0,1.6) -- (5.6,1.6);
\draw[line width=0.6mm] (0.8,2.4) -- (2.4,2.4);
\draw[line width=0.6mm] (0.8,3.2) -- (2.4,3.2);
\draw[line width=0.6mm] (1.6,4) -- (2.4,4);
\draw[line width=0.6mm] (1.6,4.8) -- (2.4,4.8);

\draw[line width=0.6mm] (0.8,2.4) -- (0.8,3.2);
\draw[line width=0.6mm] (1.6,3.2) -- (1.6,4.8);
\draw[line width=0.6mm] (2.4,1.6) -- (2.4,5.6);
\draw[line width=0.6mm] (3.2,1.6) -- (3.2,5.6);
\draw[line width=0.6mm] (4,0) -- (4,0.8);
\draw[line width=0.6mm] (4.8,0) -- (4.8,0.8);

\node at (0,1.2) {$\dots$};
\node at (5.6,1.2) {$\dots$};
\node at (2.8,5.7) {$\vdots$};

\node[draw,circle,minimum size=0.4cm, fill=blk] at (4.4,0.4) {};
\node at (3.6,1.2) {A};
\node[draw,circle,minimum size=0.4cm] at (3.6,1.2) {};
\node at (2.8,2) {B};
\node[draw,circle,minimum size=0.4cm] at (2.8,2) {};
\node[draw,circle,minimum size=0.4cm, fill=blk] at (1.2,2.8) {};
\node[draw,circle,minimum size=0.4cm] at (2,2.8) {};
\node[draw,circle,minimum size=0.4cm, fill=blk] at (2,3.6) {};
\node[draw,circle,minimum size=0.4cm, fill=blk] at (2,4.4) {};
\end{tikzpicture}
\caption{A gadget for enforcing two circles with different colors (left) and its only two solutions (right), where $\{A,B\}=\{\text{black},\text{white}\}$}
\label{fig3}
\end{figure}
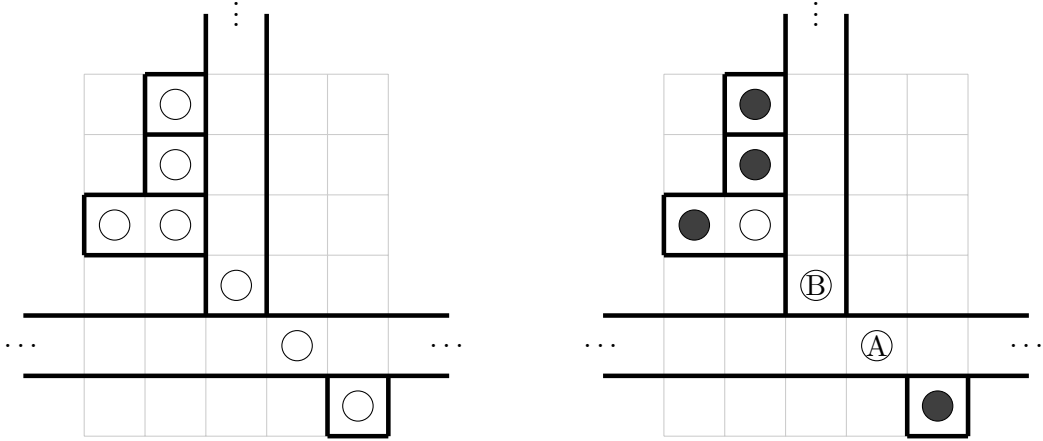

For any region with exactly two circles, the colors of these two circles in the solution are always different. However, if we want to force two circles \textit{in different regions} to have different colors in the solution, we can do so by using a gadget represented in Figure \ref{fig3}. The idea is that there are four consecutive circles arranged diagonally, with a circle at one end being forced white and at the other end being forced black. As a result, the two middle circles must have different colors in the solution (otherwise there will be three consecutive circles with the same color in the solution).

\subsection{Connecting Two Consecutive Clause Rows with Common Variables} \label{connect}
For two consecutive clause rows (e.g. clause rows for $C_2$ and $C_3$) which the corresponding clauses share a variable, we use a gadget represented in Figure \ref{fig4} (see also Figure \ref{fig5} for its solutions) to connect the two variable circles corresponding to that common variable. This gadget forces these two variable circles to have the same color in the solution, thus ensuring that the variable has the same truth value in both clauses. The idea behind this gadget is to use multiple copies of the gadget in Section \ref{diff}.

\begin{figure}
\centering
\begin{tikzpicture}
\draw[step=0.8cm,color={rgb:black,1;white,4}] (0.8,0) grid (7.2,18.4);

\draw[line width=0.6mm] (1.6,0) -- (2.4,0);
\draw[line width=0.6mm] (1.6,0.8) -- (2.4,0.8);
\draw[line width=0.6mm] (4,0.8) -- (4.8,0.8);

\draw[line width=0.6mm] (0,1.6) -- (8,1.6);
\draw[line width=0.6mm] (0,2.4) -- (8,2.4);

\draw[line width=0.6mm] (0.8,3.2) -- (2.4,3.2);
\draw[line width=0.6mm] (0.8,4) -- (2.4,4);
\draw[line width=0.6mm] (1.6,4.8) -- (2.4,4.8);
\draw[line width=0.6mm] (1.6,5.6) -- (2.4,5.6);

\draw[line width=0.6mm] (4,3.2) -- (5.6,3.2);
\draw[line width=0.6mm] (4,4) -- (4.8,4);
\draw[line width=0.6mm] (4,4.8) -- (4.8,4.8);

\draw[line width=0.6mm] (1.6,0) -- (1.6,0.8);
\draw[line width=0.6mm] (4,0.8) -- (4,1.6);
\draw[line width=0.6mm] (4.8,0.8) -- (4.8,1.6);

\draw[line width=0.6mm] (2.4,-0.8) -- (2.4,1.6);
\draw[line width=0.6mm] (3.2,-0.8) -- (3.2,1.6);
\draw[line width=0.6mm] (2.4,2.4) -- (2.4,8);
\draw[line width=0.6mm] (3.2,2.4) -- (3.2,14.4);

\draw[line width=0.6mm] (0.8,3.2) -- (0.8,4);
\draw[line width=0.6mm] (1.6,4) -- (1.6,5.6);

\draw[line width=0.6mm] (4,2.4) -- (4,4.8);
\draw[line width=0.6mm] (4.8,3.2) -- (4.8,4.8);
\draw[line width=0.6mm] (5.6,2.4) -- (5.6,3.2);

\node at (-1.5,2) {Clause Row};
\node at (0,2) {$\dots$};
\node at (8,2) {$\dots$};
\node at (2.8,-0.7) {$\vdots$};

\node[draw,circle,minimum size=0.4cm] at (2,0.4) {};
\node[draw,circle,minimum size=0.4cm] at (4.4,1.2) {};

\node[draw,circle,minimum size=0.4cm] at (2.8,1.2) {};
\node[draw,circle,minimum size=0.4cm] at (3.6,2) {};
\node[draw,circle,minimum size=0.4cm] at (2.8,2.8) {};

\node[draw,circle,minimum size=0.4cm] at (1.2,3.6) {};
\node[draw,circle,minimum size=0.4cm] at (2,3.6) {};
\node[draw,circle,minimum size=0.4cm] at (2,4.4) {};
\node[draw,circle,minimum size=0.4cm] at (2,5.2) {};

\node[draw,circle,minimum size=0.4cm] at (4.4,2.8) {};
\node[draw,circle,minimum size=0.4cm] at (5.2,2.8) {};
\node[draw,circle,minimum size=0.4cm] at (4.4,3.6) {};
\node[draw,circle,minimum size=0.4cm] at (4.4,4.4) {};

\draw[line width=0.6mm] (1.6,6.4) -- (2.4,6.4);
\draw[line width=0.6mm] (1.6,7.2) -- (2.4,7.2);
\draw[line width=0.6mm] (4,7.2) -- (4.8,7.2);

\draw[line width=0.6mm] (2.4,8) -- (7.2,8);
\draw[line width=0.6mm] (2.4,8.8) -- (7.2,8.8);

\draw[line width=0.6mm] (0.8,9.6) -- (2.4,9.6);
\draw[line width=0.6mm] (0.8,10.4) -- (2.4,10.4);
\draw[line width=0.6mm] (1.6,11.2) -- (2.4,11.2);
\draw[line width=0.6mm] (1.6,12) -- (2.4,12);

\draw[line width=0.6mm] (4,9.6) -- (5.6,9.6);
\draw[line width=0.6mm] (4,10.4) -- (4.8,10.4);
\draw[line width=0.6mm] (4,11.2) -- (4.8,11.2);

\draw[line width=0.6mm] (1.6,6.4) -- (1.6,7.2);
\draw[line width=0.6mm] (4,7.2) -- (4,8);
\draw[line width=0.6mm] (4.8,7.2) -- (4.8,8);

\draw[line width=0.6mm] (7.2,8) -- (7.2,8.8);

\draw[line width=0.6mm] (0.8,9.6) -- (0.8,10.4);
\draw[line width=0.6mm] (1.6,10.4) -- (1.6,12);

\draw[line width=0.6mm] (4,8.8) -- (4,11.2);
\draw[line width=0.6mm] (4.8,9.6) -- (4.8,11.2);
\draw[line width=0.6mm] (5.6,8.8) -- (5.6,9.6);

\node[draw,circle,minimum size=0.4cm] at (2,6.8) {};
\node[draw,circle,minimum size=0.4cm] at (4.4,7.6) {};

\node[draw,circle,minimum size=0.4cm] at (2.8,7.6) {};
\node[draw,circle,minimum size=0.4cm] at (3.6,8.4) {};
\node[draw,circle,minimum size=0.4cm] at (2.8,9.2) {};
\node[draw,circle,minimum size=0.4cm] at (6.8,8.4) {};

\node[draw,circle,minimum size=0.4cm] at (1.2,10) {};
\node[draw,circle,minimum size=0.4cm] at (2,10) {};
\node[draw,circle,minimum size=0.4cm] at (2,10.8) {};
\node[draw,circle,minimum size=0.4cm] at (2,11.6) {};

\node[draw,circle,minimum size=0.4cm] at (4.4,9.2) {};
\node[draw,circle,minimum size=0.4cm] at (5.2,9.2) {};
\node[draw,circle,minimum size=0.4cm] at (4.4,10) {};
\node[draw,circle,minimum size=0.4cm] at (4.4,10.8) {};

\draw[line width=0.6mm] (1.6,12.8) -- (2.4,12.8);
\draw[line width=0.6mm] (1.6,13.6) -- (2.4,13.6);
\draw[line width=0.6mm] (4,13.6) -- (4.8,13.6);

\draw[line width=0.6mm] (0,14.4) -- (8,14.4);
\draw[line width=0.6mm] (0,15.2) -- (8,15.2);

\draw[line width=0.6mm] (0.8,16) -- (2.4,16);
\draw[line width=0.6mm] (0.8,16.8) -- (2.4,16.8);
\draw[line width=0.6mm] (1.6,17.6) -- (2.4,17.6);
\draw[line width=0.6mm] (1.6,18.4) -- (2.4,18.4);

\draw[line width=0.6mm] (4,16) -- (5.6,16);
\draw[line width=0.6mm] (4,16.8) -- (4.8,16.8);
\draw[line width=0.6mm] (4,17.6) -- (4.8,17.6);

\draw[line width=0.6mm] (1.6,12.8) -- (1.6,13.6);
\draw[line width=0.6mm] (4,13.6) -- (4,14.4);
\draw[line width=0.6mm] (4.8,13.6) -- (4.8,14.4);

\draw[line width=0.6mm] (2.4,8.8) -- (2.4,14.4);
\draw[line width=0.6mm] (2.4,15.2) -- (2.4,19.2);
\draw[line width=0.6mm] (3.2,15.2) -- (3.2,19.2);

\draw[line width=0.6mm] (0.8,16) -- (0.8,16.8);
\draw[line width=0.6mm] (1.6,16.8) -- (1.6,18.4);

\draw[line width=0.6mm] (4,15.2) -- (4,17.6);
\draw[line width=0.6mm] (4.8,16) -- (4.8,17.6);
\draw[line width=0.6mm] (5.6,15.2) -- (5.6,16);

\node at (-1.5,14.8) {Clause Row};
\node at (0,14.8) {$\dots$};
\node at (8,14.8) {$\dots$};
\node at (2.8,19.3) {$\vdots$};
\node[rotate=90] at (3.6,18) {Variable Column};

\node[draw,circle,minimum size=0.4cm] at (2,13.2) {};
\node[draw,circle,minimum size=0.4cm] at (4.4,14) {};

\node[draw,circle,minimum size=0.4cm] at (2.8,14) {};
\node[draw,circle,minimum size=0.4cm] at (3.6,14.8) {};
\node[draw,circle,minimum size=0.4cm] at (2.8,15.6) {};

\node[draw,circle,minimum size=0.4cm] at (1.2,16.4) {};
\node[draw,circle,minimum size=0.4cm] at (2,16.4) {};
\node[draw,circle,minimum size=0.4cm] at (2,17.2) {};
\node[draw,circle,minimum size=0.4cm] at (2,18) {};

\node[draw,circle,minimum size=0.4cm] at (4.4,15.6) {};
\node[draw,circle,minimum size=0.4cm] at (5.2,15.6) {};
\node[draw,circle,minimum size=0.4cm] at (4.4,16.4) {};
\node[draw,circle,minimum size=0.4cm] at (4.4,17.2) {};
\end{tikzpicture}
\caption{A gadget connecting two consecutive clause rows, forcing the two variable circles to have the same color}
\label{fig4}
\end{figure}
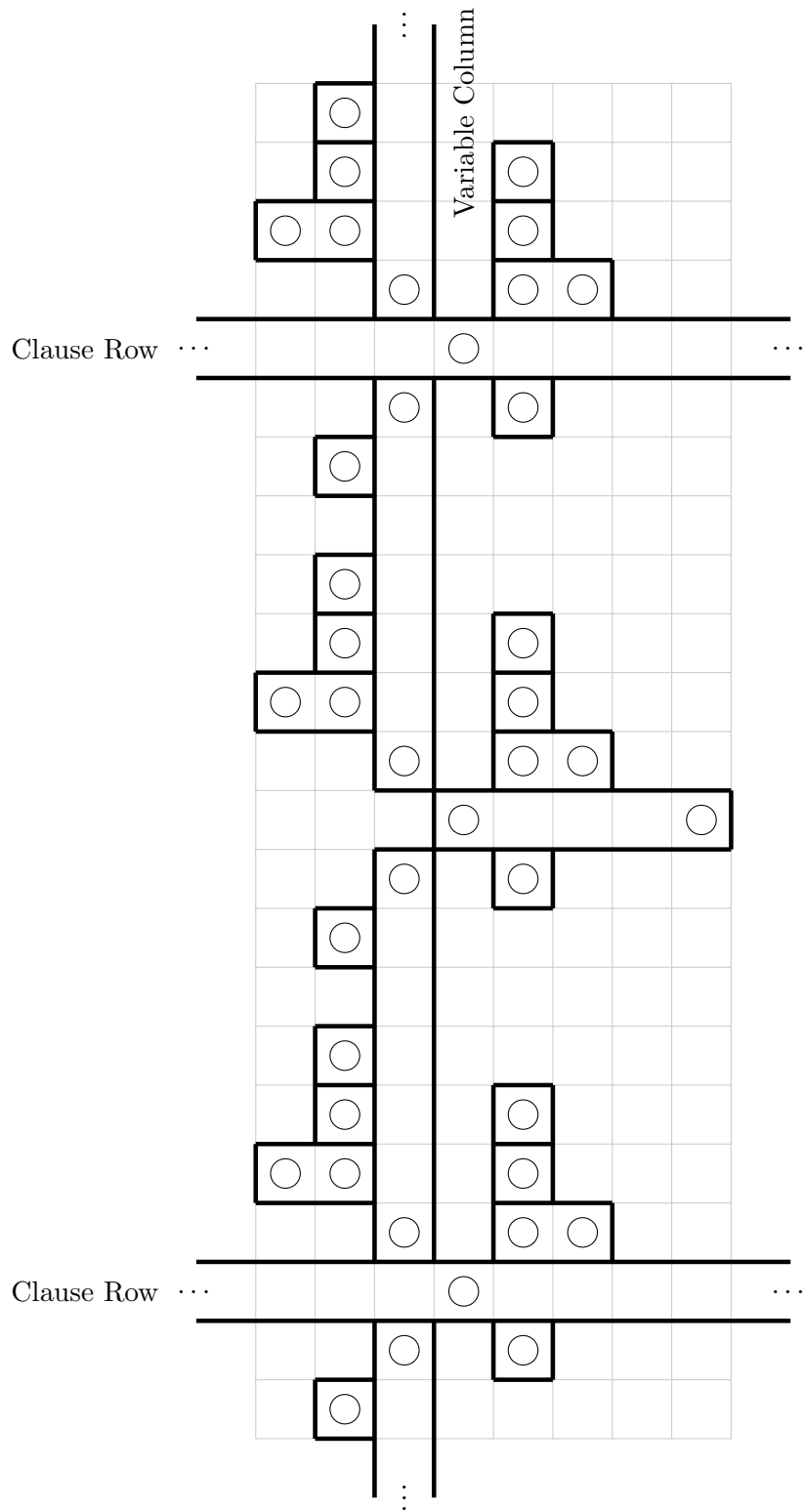

\begin{figure}
\centering
\begin{tikzpicture}
\draw[step=0.8cm,color={rgb:black,1;white,4}] (0.8,0) grid (7.2,18.4);

\draw[line width=0.6mm] (1.6,0) -- (2.4,0);
\draw[line width=0.6mm] (1.6,0.8) -- (2.4,0.8);
\draw[line width=0.6mm] (4,0.8) -- (4.8,0.8);

\draw[line width=0.6mm] (0,1.6) -- (8,1.6);
\draw[line width=0.6mm] (0,2.4) -- (8,2.4);

\draw[line width=0.6mm] (0.8,3.2) -- (2.4,3.2);
\draw[line width=0.6mm] (0.8,4) -- (2.4,4);
\draw[line width=0.6mm] (1.6,4.8) -- (2.4,4.8);
\draw[line width=0.6mm] (1.6,5.6) -- (2.4,5.6);

\draw[line width=0.6mm] (4,3.2) -- (5.6,3.2);
\draw[line width=0.6mm] (4,4) -- (4.8,4);
\draw[line width=0.6mm] (4,4.8) -- (4.8,4.8);

\draw[line width=0.6mm] (1.6,0) -- (1.6,0.8);
\draw[line width=0.6mm] (4,0.8) -- (4,1.6);
\draw[line width=0.6mm] (4.8,0.8) -- (4.8,1.6);

\draw[line width=0.6mm] (2.4,-0.8) -- (2.4,1.6);
\draw[line width=0.6mm] (3.2,-0.8) -- (3.2,1.6);
\draw[line width=0.6mm] (2.4,2.4) -- (2.4,8);
\draw[line width=0.6mm] (3.2,2.4) -- (3.2,14.4);

\draw[line width=0.6mm] (0.8,3.2) -- (0.8,4);
\draw[line width=0.6mm] (1.6,4) -- (1.6,5.6);

\draw[line width=0.6mm] (4,2.4) -- (4,4.8);
\draw[line width=0.6mm] (4.8,3.2) -- (4.8,4.8);
\draw[line width=0.6mm] (5.6,2.4) -- (5.6,3.2);

\node at (-1.5,2) {Clause Row};
\node at (0,2) {$\dots$};
\node at (8,2) {$\dots$};
\node at (2.8,-0.7) {$\vdots$};

\node[draw,circle,minimum size=0.4cm, fill=blk] at (2,0.4) {};
\node[draw,circle,minimum size=0.4cm, fill=blk] at (4.4,1.2) {};

\node at (2.8,1.2) {B};
\node[draw,circle,minimum size=0.4cm] at (2.8,1.2) {};
\node at (3.6,2) {A};
\node[draw,circle,minimum size=0.4cm] at (3.6,2) {};
\node at (2.8,2.8) {B};
\node[draw,circle,minimum size=0.4cm] at (2.8,2.8) {};

\node[draw,circle,minimum size=0.4cm, fill=blk] at (1.2,3.6) {};
\node[draw,circle,minimum size=0.4cm] at (2,3.6) {};
\node[draw,circle,minimum size=0.4cm, fill=blk] at (2,4.4) {};
\node[draw,circle,minimum size=0.4cm, fill=blk] at (2,5.2) {};

\node[draw,circle,minimum size=0.4cm] at (4.4,2.8) {};
\node[draw,circle,minimum size=0.4cm, fill=blk] at (5.2,2.8) {};
\node[draw,circle,minimum size=0.4cm, fill=blk] at (4.4,3.6) {};
\node[draw,circle,minimum size=0.4cm, fill=blk] at (4.4,4.4) {};

\draw[line width=0.6mm] (1.6,6.4) -- (2.4,6.4);
\draw[line width=0.6mm] (1.6,7.2) -- (2.4,7.2);
\draw[line width=0.6mm] (4,7.2) -- (4.8,7.2);

\draw[line width=0.6mm] (2.4,8) -- (7.2,8);
\draw[line width=0.6mm] (2.4,8.8) -- (7.2,8.8);

\draw[line width=0.6mm] (0.8,9.6) -- (2.4,9.6);
\draw[line width=0.6mm] (0.8,10.4) -- (2.4,10.4);
\draw[line width=0.6mm] (1.6,11.2) -- (2.4,11.2);
\draw[line width=0.6mm] (1.6,12) -- (2.4,12);

\draw[line width=0.6mm] (4,9.6) -- (5.6,9.6);
\draw[line width=0.6mm] (4,10.4) -- (4.8,10.4);
\draw[line width=0.6mm] (4,11.2) -- (4.8,11.2);

\draw[line width=0.6mm] (1.6,6.4) -- (1.6,7.2);
\draw[line width=0.6mm] (4,7.2) -- (4,8);
\draw[line width=0.6mm] (4.8,7.2) -- (4.8,8);

\draw[line width=0.6mm] (7.2,8) -- (7.2,8.8);

\draw[line width=0.6mm] (0.8,9.6) -- (0.8,10.4);
\draw[line width=0.6mm] (1.6,10.4) -- (1.6,12);

\draw[line width=0.6mm] (4,8.8) -- (4,11.2);
\draw[line width=0.6mm] (4.8,9.6) -- (4.8,11.2);
\draw[line width=0.6mm] (5.6,8.8) -- (5.6,9.6);

\node[draw,circle,minimum size=0.4cm, fill=blk] at (2,6.8) {};
\node[draw,circle,minimum size=0.4cm, fill=blk] at (4.4,7.6) {};

\node at (2.8,7.6) {A};
\node[draw,circle,minimum size=0.4cm] at (2.8,7.6) {};
\node at (3.6,8.4) {B};
\node[draw,circle,minimum size=0.4cm] at (3.6,8.4) {};
\node at (2.8,9.2) {A};
\node[draw,circle,minimum size=0.4cm] at (2.8,9.2) {};
\node at (6.8,8.4) {A};
\node[draw,circle,minimum size=0.4cm] at (6.8,8.4) {};

\node[draw,circle,minimum size=0.4cm, fill=blk] at (1.2,10) {};
\node[draw,circle,minimum size=0.4cm] at (2,10) {};
\node[draw,circle,minimum size=0.4cm, fill=blk] at (2,10.8) {};
\node[draw,circle,minimum size=0.4cm, fill=blk] at (2,11.6) {};

\node[draw,circle,minimum size=0.4cm] at (4.4,9.2) {};
\node[draw,circle,minimum size=0.4cm, fill=blk] at (5.2,9.2) {};
\node[draw,circle,minimum size=0.4cm, fill=blk] at (4.4,10) {};
\node[draw,circle,minimum size=0.4cm, fill=blk] at (4.4,10.8) {};

\draw[line width=0.6mm] (1.6,12.8) -- (2.4,12.8);
\draw[line width=0.6mm] (1.6,13.6) -- (2.4,13.6);
\draw[line width=0.6mm] (4,13.6) -- (4.8,13.6);

\draw[line width=0.6mm] (0,14.4) -- (8,14.4);
\draw[line width=0.6mm] (0,15.2) -- (8,15.2);

\draw[line width=0.6mm] (0.8,16) -- (2.4,16);
\draw[line width=0.6mm] (0.8,16.8) -- (2.4,16.8);
\draw[line width=0.6mm] (1.6,17.6) -- (2.4,17.6);
\draw[line width=0.6mm] (1.6,18.4) -- (2.4,18.4);

\draw[line width=0.6mm] (4,16) -- (5.6,16);
\draw[line width=0.6mm] (4,16.8) -- (4.8,16.8);
\draw[line width=0.6mm] (4,17.6) -- (4.8,17.6);

\draw[line width=0.6mm] (1.6,12.8) -- (1.6,13.6);
\draw[line width=0.6mm] (4,13.6) -- (4,14.4);
\draw[line width=0.6mm] (4.8,13.6) -- (4.8,14.4);

\draw[line width=0.6mm] (2.4,8.8) -- (2.4,14.4);
\draw[line width=0.6mm] (2.4,15.2) -- (2.4,19.2);
\draw[line width=0.6mm] (3.2,15.2) -- (3.2,19.2);

\draw[line width=0.6mm] (0.8,16) -- (0.8,16.8);
\draw[line width=0.6mm] (1.6,16.8) -- (1.6,18.4);

\draw[line width=0.6mm] (4,15.2) -- (4,17.6);
\draw[line width=0.6mm] (4.8,16) -- (4.8,17.6);
\draw[line width=0.6mm] (5.6,15.2) -- (5.6,16);

\node at (-1.5,14.8) {Clause Row};
\node at (0,14.8) {$\dots$};
\node at (8,14.8) {$\dots$};
\node at (2.8,19.3) {$\vdots$};
\node[rotate=90] at (3.6,18) {Variable Column};

\node[draw,circle,minimum size=0.4cm, fill=blk] at (2,13.2) {};
\node[draw,circle,minimum size=0.4cm, fill=blk] at (4.4,14) {};

\node at (2.8,14) {B};
\node[draw,circle,minimum size=0.4cm] at (2.8,14) {};
\node at (3.6,14.8) {A};
\node[draw,circle,minimum size=0.4cm] at (3.6,14.8) {};
\node at (2.8,15.6) {B};
\node[draw,circle,minimum size=0.4cm] at (2.8,15.6) {};

\node[draw,circle,minimum size=0.4cm, fill=blk] at (1.2,16.4) {};
\node[draw,circle,minimum size=0.4cm] at (2,16.4) {};
\node[draw,circle,minimum size=0.4cm, fill=blk] at (2,17.2) {};
\node[draw,circle,minimum size=0.4cm, fill=blk] at (2,18) {};

\node[draw,circle,minimum size=0.4cm] at (4.4,15.6) {};
\node[draw,circle,minimum size=0.4cm, fill=blk] at (5.2,15.6) {};
\node[draw,circle,minimum size=0.4cm, fill=blk] at (4.4,16.4) {};
\node[draw,circle,minimum size=0.4cm, fill=blk] at (4.4,17.2) {};
\end{tikzpicture}
\caption{The only two solutions of the puzzle in Figure \ref{fig4}, where $\{A,B\}=\{\text{black},\text{white}\}$}
\label{fig5}
\end{figure}
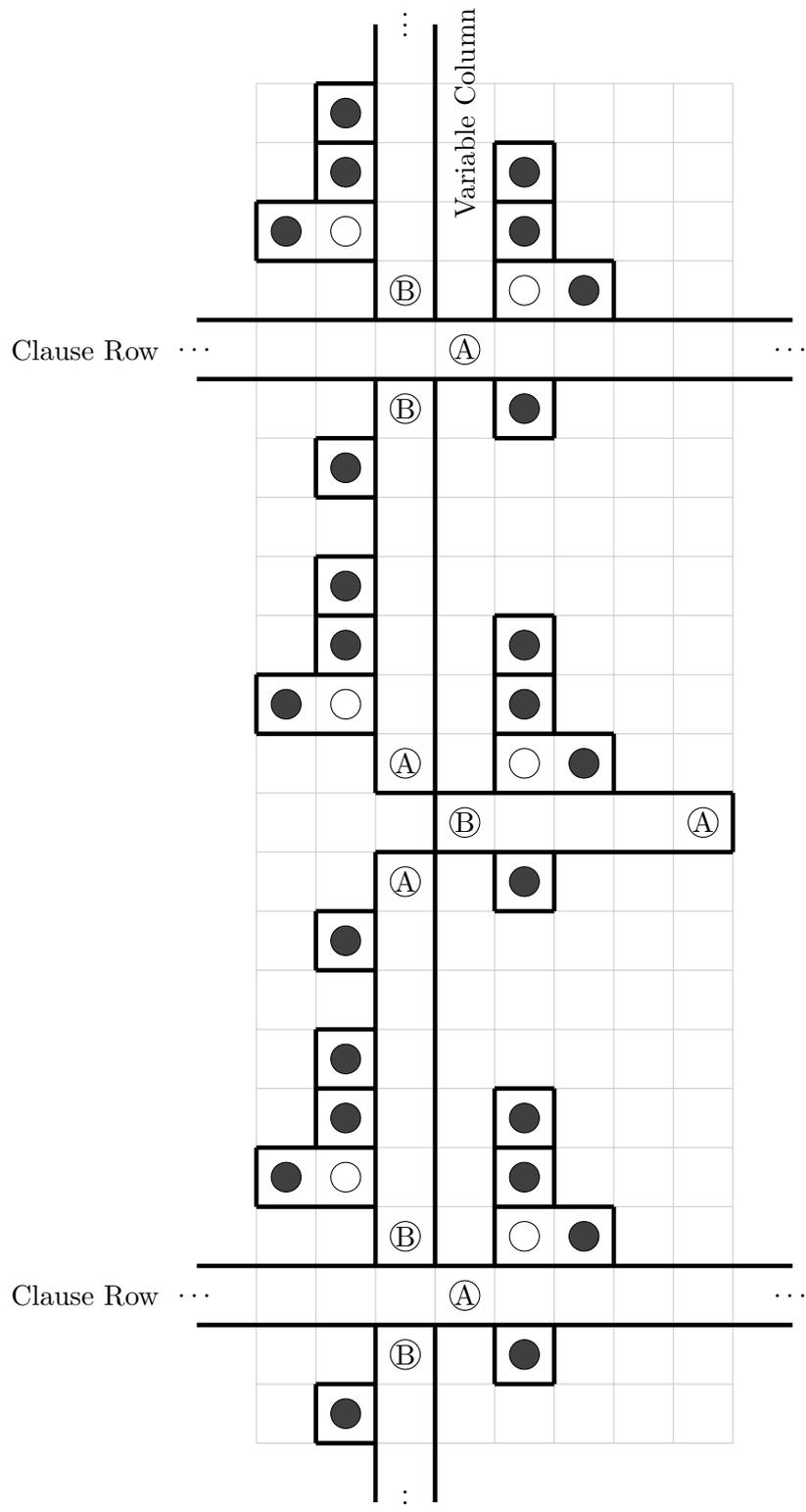

\subsection{Skipping a Clause Row}
As the gadget in the Section \ref{connect} can only connect consecutive clause rows, we also provide a method to skip a clause row that does not contain the given variable. For example, if $x_1$ appears in $C_2$ and $C_4$, but not in $C_3$, we need to connect the clause rows of $C_2$ and $C_4$ while skipping the one of $C_3$\footnote{In fact, this gadget is unnecessary if we instead construct a reduction from the planar positive rectilinear 1-in-3-SAT problem, which is also NP-complete \cite{planar}, where the clause rows can be arranged such that we do not need to skip a clause row. However, we include this gadget for the sake of completeness.}.

We can do so by using a gadget represented in Figure \ref{fig6} (see also Figure \ref{fig7} for its solutions). The idea behind this gadget is that we put a forced white circle as a variable circle in the clause row we want to skip, so that the color constraint for other circles in that row will not be affected.

\begin{figure}
\centering
\begin{tikzpicture}
\draw[step=0.8cm,color={rgb:black,1;white,4}] (0.8,0) grid (6.4,16);

\draw[line width=0.6mm] (1.6,0) -- (2.4,0);
\draw[line width=0.6mm] (1.6,0.8) -- (2.4,0.8);
\draw[line width=0.6mm] (4,0.8) -- (4.8,0.8);

\draw[line width=0.6mm] (0,1.6) -- (7.2,1.6);
\draw[line width=0.6mm] (0,2.4) -- (7.2,2.4);

\draw[line width=0.6mm] (0.8,3.2) -- (2.4,3.2);
\draw[line width=0.6mm] (0.8,4) -- (2.4,4);
\draw[line width=0.6mm] (1.6,4.8) -- (2.4,4.8);
\draw[line width=0.6mm] (1.6,5.6) -- (2.4,5.6);

\draw[line width=0.6mm] (4,3.2) -- (5.6,3.2);
\draw[line width=0.6mm] (4,4) -- (4.8,4);
\draw[line width=0.6mm] (4,4.8) -- (4.8,4.8);

\draw[line width=0.6mm] (1.6,0) -- (1.6,0.8);
\draw[line width=0.6mm] (4,0.8) -- (4,1.6);
\draw[line width=0.6mm] (4.8,0.8) -- (4.8,1.6);

\draw[line width=0.6mm] (2.4,-0.8) -- (2.4,1.6);
\draw[line width=0.6mm] (3.2,-0.8) -- (3.2,1.6);
\draw[line width=0.6mm] (2.4,2.4) -- (2.4,7.2);
\draw[line width=0.6mm] (3.2,2.4) -- (3.2,6.4);

\draw[line width=0.6mm] (0.8,3.2) -- (0.8,4);
\draw[line width=0.6mm] (1.6,4) -- (1.6,5.6);

\draw[line width=0.6mm] (4,2.4) -- (4,4.8);
\draw[line width=0.6mm] (4.8,3.2) -- (4.8,4.8);
\draw[line width=0.6mm] (5.6,2.4) -- (5.6,3.2);

\node at (-1.5,2) {Clause Row};
\node at (0,2) {$\dots$};
\node at (8,2) {$\dots$};
\node at (2.8,-0.7) {$\vdots$};

\node[draw,circle,minimum size=0.4cm] at (2,0.4) {};
\node[draw,circle,minimum size=0.4cm] at (4.4,1.2) {};

\node[draw,circle,minimum size=0.4cm] at (2.8,1.2) {};
\node[draw,circle,minimum size=0.4cm] at (3.6,2) {};
\node[draw,circle,minimum size=0.4cm] at (2.8,2.8) {};

\node[draw,circle,minimum size=0.4cm] at (1.2,3.6) {};
\node[draw,circle,minimum size=0.4cm] at (2,3.6) {};
\node[draw,circle,minimum size=0.4cm] at (2,4.4) {};
\node[draw,circle,minimum size=0.4cm] at (2,5.2) {};

\node[draw,circle,minimum size=0.4cm] at (4.4,2.8) {};
\node[draw,circle,minimum size=0.4cm] at (5.2,2.8) {};
\node[draw,circle,minimum size=0.4cm] at (4.4,3.6) {};
\node[draw,circle,minimum size=0.4cm] at (4.4,4.4) {};

\draw[line width=0.6mm] (3.2,6.4) -- (5.6,6.4);
\draw[line width=0.6mm] (2.4,7.2) -- (4.8,7.2);

\draw[line width=0.6mm] (5.6,8) -- (6.4,8);
\draw[line width=0.6mm] (5.6,8.8) -- (6.4,8.8);

\draw[line width=0.6mm] (2.4,9.6) -- (5.6,9.6);
\draw[line width=0.6mm] (0,10.4) -- (7.2,10.4);
\draw[line width=0.6mm] (0,11.2) -- (7.2,11.2);
\draw[line width=0.6mm] (1.6,12) -- (4.8,12);

\draw[line width=0.6mm] (4.8,7.2) -- (4.8,10.4);
\draw[line width=0.6mm] (5.6,6.4) -- (5.6,9.6);

\draw[line width=0.6mm] (6.4,8) -- (6.4,8.8);

\draw[line width=0.6mm] (2.4,9.6) -- (2.4,10.4);
\draw[line width=0.6mm] (3.2,9.6) -- (3.2,10.4);
\draw[line width=0.6mm] (4,11.2) -- (4,12);
\draw[line width=0.6mm] (4.8,11.2) -- (4.8,12);

\node at (-1.5,11.2) {Clause Row};
\node at (-1.5,10.8) {we want};
\node at (-1.5,10.4) {to skip};
\node at (0,10.8) {$\dots$};
\node at (8,10.8) {$\dots$};

\node[draw,circle,minimum size=0.4cm] at (6,8.4) {};
\node[draw,circle,minimum size=0.4cm] at (5.2,9.2) {};

\node[draw,circle,minimum size=0.4cm] at (2.8,10) {};
\node[draw,circle,minimum size=0.4cm] at (3.6,10) {};
\node[draw,circle,minimum size=0.4cm] at (4.4,10) {};

\node[draw,circle,minimum size=0.4cm] at (3.6,10.8) {};

\node[draw,circle,minimum size=0.4cm] at (2.8,11.6) {};
\node[draw,circle,minimum size=0.4cm] at (3.6,11.6) {};
\node[draw,circle,minimum size=0.4cm] at (4.4,11.6) {};

\draw[line width=0.6mm] (0.8,12.8) -- (1.6,12.8);
\draw[line width=0.6mm] (0.8,13.6) -- (1.6,13.6);
\draw[line width=0.6mm] (2.4,14.4) -- (3.2,14.4);
\draw[line width=0.6mm] (1.6,15.2) -- (2.4,15.2);

\draw[line width=0.6mm] (0.8,12.8) -- (0.8,13.6);
\draw[line width=0.6mm] (1.6,12) -- (1.6,15.2);
\draw[line width=0.6mm] (2.4,11.2) -- (2.4,14.4);
\draw[line width=0.6mm] (2.4,15.2) -- (2.4,16.8);
\draw[line width=0.6mm] (3.2,14.4) -- (3.2,16.8);

\node at (2.8,16.9) {$\vdots$};
\node[rotate=90] at (3.6,17.6) {Variable Column};

\node[draw,circle,minimum size=0.4cm] at (2,12.4) {};
\node[draw,circle,minimum size=0.4cm] at (1.2,13.2) {};
\end{tikzpicture}
\caption{A gadget to skip a clause row, starting from the lower clause row and skipping the upper clause row to connect to some clause row beyond it}
\label{fig6}
\end{figure}

\begin{figure}
\centering
\begin{tikzpicture}
\draw[step=0.8cm,color={rgb:black,1;white,4}] (0.8,0) grid (6.4,16);

\draw[line width=0.6mm] (1.6,0) -- (2.4,0);
\draw[line width=0.6mm] (1.6,0.8) -- (2.4,0.8);
\draw[line width=0.6mm] (4,0.8) -- (4.8,0.8);

\draw[line width=0.6mm] (0,1.6) -- (7.2,1.6);
\draw[line width=0.6mm] (0,2.4) -- (7.2,2.4);

\draw[line width=0.6mm] (0.8,3.2) -- (2.4,3.2);
\draw[line width=0.6mm] (0.8,4) -- (2.4,4);
\draw[line width=0.6mm] (1.6,4.8) -- (2.4,4.8);
\draw[line width=0.6mm] (1.6,5.6) -- (2.4,5.6);

\draw[line width=0.6mm] (4,3.2) -- (5.6,3.2);
\draw[line width=0.6mm] (4,4) -- (4.8,4);
\draw[line width=0.6mm] (4,4.8) -- (4.8,4.8);

\draw[line width=0.6mm] (1.6,0) -- (1.6,0.8);
\draw[line width=0.6mm] (4,0.8) -- (4,1.6);
\draw[line width=0.6mm] (4.8,0.8) -- (4.8,1.6);

\draw[line width=0.6mm] (2.4,-0.8) -- (2.4,1.6);
\draw[line width=0.6mm] (3.2,-0.8) -- (3.2,1.6);
\draw[line width=0.6mm] (2.4,2.4) -- (2.4,7.2);
\draw[line width=0.6mm] (3.2,2.4) -- (3.2,6.4);

\draw[line width=0.6mm] (0.8,3.2) -- (0.8,4);
\draw[line width=0.6mm] (1.6,4) -- (1.6,5.6);

\draw[line width=0.6mm] (4,2.4) -- (4,4.8);
\draw[line width=0.6mm] (4.8,3.2) -- (4.8,4.8);
\draw[line width=0.6mm] (5.6,2.4) -- (5.6,3.2);

\node at (-1.5,2) {Clause Row};
\node at (0,2) {$\dots$};
\node at (8,2) {$\dots$};
\node at (2.8,-0.7) {$\vdots$};

\node[draw,circle,minimum size=0.4cm, fill=blk] at (2,0.4) {};
\node[draw,circle,minimum size=0.4cm, fill=blk] at (4.4,1.2) {};

\node at (2.8,1.2) {B};
\node[draw,circle,minimum size=0.4cm] at (2.8,1.2) {};
\node at (3.6,2) {A};
\node[draw,circle,minimum size=0.4cm] at (3.6,2) {};
\node at (2.8,2.8) {B};
\node[draw,circle,minimum size=0.4cm] at (2.8,2.8) {};

\node[draw,circle,minimum size=0.4cm, fill=blk] at (1.2,3.6) {};
\node[draw,circle,minimum size=0.4cm] at (2,3.6) {};
\node[draw,circle,minimum size=0.4cm, fill=blk] at (2,4.4) {};
\node[draw,circle,minimum size=0.4cm, fill=blk] at (2,5.2) {};

\node[draw,circle,minimum size=0.4cm] at (4.4,2.8) {};
\node[draw,circle,minimum size=0.4cm, fill=blk] at (5.2,2.8) {};
\node[draw,circle,minimum size=0.4cm, fill=blk] at (4.4,3.6) {};
\node[draw,circle,minimum size=0.4cm, fill=blk] at (4.4,4.4) {};

\draw[line width=0.6mm] (3.2,6.4) -- (5.6,6.4);
\draw[line width=0.6mm] (2.4,7.2) -- (4.8,7.2);

\draw[line width=0.6mm] (5.6,8) -- (6.4,8);
\draw[line width=0.6mm] (5.6,8.8) -- (6.4,8.8);

\draw[line width=0.6mm] (2.4,9.6) -- (5.6,9.6);
\draw[line width=0.6mm] (0,10.4) -- (7.2,10.4);
\draw[line width=0.6mm] (0,11.2) -- (7.2,11.2);
\draw[line width=0.6mm] (1.6,12) -- (4.8,12);

\draw[line width=0.6mm] (4.8,7.2) -- (4.8,10.4);
\draw[line width=0.6mm] (5.6,6.4) -- (5.6,9.6);

\draw[line width=0.6mm] (6.4,8) -- (6.4,8.8);

\draw[line width=0.6mm] (2.4,9.6) -- (2.4,10.4);
\draw[line width=0.6mm] (3.2,9.6) -- (3.2,10.4);
\draw[line width=0.6mm] (4,11.2) -- (4,12);
\draw[line width=0.6mm] (4.8,11.2) -- (4.8,12);

\node at (-1.5,11.2) {Clause Row};
\node at (-1.5,10.8) {we want};
\node at (-1.5,10.4) {to skip};
\node at (0,10.8) {$\dots$};
\node at (8,10.8) {$\dots$};

\node[draw,circle,minimum size=0.4cm, fill=blk] at (6,8.4) {};
\node at (5.2,9.2) {A};
\node[draw,circle,minimum size=0.4cm] at (5.2,9.2) {};

\node[draw,circle,minimum size=0.4cm, fill=blk] at (2.8,10) {};
\node at (3.6,10) {A};
\node[draw,circle,minimum size=0.4cm] at (3.6,10) {};
\node at (4.4,10) {B};
\node[draw,circle,minimum size=0.4cm] at (4.4,10) {};

\node[draw,circle,minimum size=0.4cm] at (3.6,10.8) {};

\node at (2.8,11.6) {A};
\node[draw,circle,minimum size=0.4cm] at (2.8,11.6) {};
\node at (3.6,11.6) {B};
\node[draw,circle,minimum size=0.4cm] at (3.6,11.6) {};
\node[draw,circle,minimum size=0.4cm, fill=blk] at (4.4,11.6) {};

\draw[line width=0.6mm] (0.8,12.8) -- (1.6,12.8);
\draw[line width=0.6mm] (0.8,13.6) -- (1.6,13.6);
\draw[line width=0.6mm] (2.4,14.4) -- (3.2,14.4);
\draw[line width=0.6mm] (1.6,15.2) -- (2.4,15.2);

\draw[line width=0.6mm] (0.8,12.8) -- (0.8,13.6);
\draw[line width=0.6mm] (1.6,12) -- (1.6,15.2);
\draw[line width=0.6mm] (2.4,11.2) -- (2.4,14.4);
\draw[line width=0.6mm] (2.4,15.2) -- (2.4,16.8);
\draw[line width=0.6mm] (3.2,14.4) -- (3.2,16.8);

\node at (2.8,16.9) {$\vdots$};
\node[rotate=90] at (3.6,17.6) {Variable Column};

\node at (2,12.4) {B};
\node[draw,circle,minimum size=0.4cm] at (2,12.4) {};
\node[draw,circle,minimum size=0.4cm, fill=blk] at (1.2,13.2) {};
\end{tikzpicture}
\caption{The only two solutions of the puzzle in Figure \ref{fig6}, where $\{A,B\}=\{\text{black},\text{white}\}$}
\label{fig7}
\end{figure}

\subsection{Filling Empty Area}
We can simply make each connected empty area into one region, with one circle placed inside it, not touching any boundary. That circle is forced to be black without affecting other regions.

Recall that we interpret a black (resp. white) circle in a Nondango solution as a true (resp. false) literal. We can see that the Nondango puzzle we construct has a solution if and only if the original 1-in-3-SAT+ problem is satisfiable. As the reduction is clearly parsimonious, we can conclude that deciding solvability of a given Nondango puzzle is NP-complete.
\newpage


\begin{thebibliography}{99}
	\bibitem{numberlink} A. Adcock, E.D. Demaine, M.L. Demaine, M.P. O'Brien, F. Reidl, F.S. Villaamil and B.D. Sullivan. Zig-Zag Numberlink is NP-Complete. \textit{Journal of Information Processing}, 23(3): 239--245 (2015).
	\bibitem{tatamibari} A. Adler, J. Bosboom, E.D. Demaine, M.L. Demaine, Q.C. Liu and J. Lynch. Tatamibari Is NP-Complete. In \textit{Proceedings of the 10th International Conference on Fun with Algorithms (FUN)}, pp. 1:1--1:24 (2020).
	\bibitem{stone} A. Allen and A. Williams. Sto-Stone is NP-Complete. In \textit{Proceedings of the 30th Canadian Conference on Computational Geometry (CCCG)}, pp. 28--34 (2018).
	\bibitem{bridges} D. Andersson. Hashiwokakero is NP-complete. \textit{Information Processing Letters}, 109(9): 1145--1146 (2009).
	\bibitem{goishi} D. Andersson. HIROIMONO Is NP-Complete. In \textit{Proceedings of the 4th International Conference on Fun with Algorithms (FUN)}, pp. 30--39 (2007).
	\bibitem{lits} M. Biro and C. Schmidt. Computational complexity and bounds for Norinori and LITS. In \textit{Proceedings of the 33rd European Workshop on Computational Geometry (EuroCG)}, pp. 29--32 (2017).
	\bibitem{yinyang} E.D. Demaine, J. Lynch, M. Rudoy and Y. Uno. Yin-Yang Puzzles are NP-complete. In \textit{Proceedings of the 33rd Canadian Conference on Computational Geometry (CCCG)}, pp. 97--106 (2021).
	\bibitem{shakashaka} E.D. Demaine, Y. Okamoto, R. Uehara and Y. Uno. Computational Complexity and an Integer Programming Model of Shakashaka. \textit{IEICE Trans. Fundamentals}, 97.A(6): 1213--1219 (2014).
	\bibitem{roma} K. Goergen, H. Fernau, E. Oest and P. Wolf. All Paths Lead to Rome. In \textit{Proceedings of the 24th Japan Conference on Discrete and Computational Geometry, Graphs, and Games (JCDC$G^3$)}, pp. 38--39 (2022).
	\bibitem{nurikabe} M. Holzer, A. Klein and M. Kutrib. On The NP-Completeness of The Nurikabe Pencil Puzzle and Variants Thereof. In \textit{Proceedings of the 3rd International Conference on Fun with Algorithms (FUN)}, pp. 77--89 (2004).
	\bibitem{heyawake} M. Holzer and O. Ruepp. The Troubles of Interior Design–A Complexity Analysis of the Game Heyawake. In \textit{Proceedings of the 4th International Conference on Fun with Algorithms (FUN)}, pp. 198--212 (2007).
	\bibitem{yosenabe} C. Iwamoto. Yosenabe is NP-complete. \textit{Journal of Information Processing}, 22(1): 40--43 (2014).
	\bibitem{usowan} C. Iwamoto and M. Haruishi. Computational Complexity of Usowan Puzzles. \textit{IEICE Trans. Fundamentals}, 101.A(9): 1537--1540 (2018).
	\bibitem{makaro} C. Iwamoto, M. Haruishi and T. Ibusuki. Herugolf and Makaro are NP-complete. In \textit{Proceedings of the 9th International Conference on Fun with Algorithms (FUN)}, pp. 24:1--24:11 (2018).
	\bibitem{dosun} C. Iwamoto and T. Ibusuki. Dosun-Fuwari is NP-complete. \textit{Journal of Information Processing}, 26: 358--361 (2018).
	\bibitem{juosan} C. Iwamoto and T. Ibusuki. Polynomial-Time Reductions from 3SAT to Kurotto and Juosan Puzzles. \textit{IEICE Trans. Inf. \& Syst.}, 103.D(3): 500--505 (2020).
	\bibitem{nurimisaki} C. Iwamoto and T. Ide. Computational Complexity of Nurimisaki and Sashigane. \textit{IEICE Trans. Fundamentals}, 103.A(10): 1183--1192 (2020).
	\bibitem{fivecells} C. Iwamoto and T. Ide. Five Cells and Tilepaint are NP-Complete. \textit{IEICE Trans. Inf. \& Syst.}, 105.D(3): 508--516 (2022).
	\bibitem{moon} C. Iwamoto and T. Ide. Moon-or-Sun, Nagareru, and Nurimeizu are NP-Complete. \textit{IEICE Trans. Fundamentals}, 105.A(9): 1187--1194 (2022).
	\bibitem{janko} A. Janko and O. Janko. Nondango. \url{https://www.janko.at/Raetsel/Nondango/index.htm}
	\bibitem{kurodoko} J. K\"olker. Kurodoko is NP-Complete. \textit{Journal of Information Processing}, 20(3): 694--706 (2012).
	\bibitem{planar} W. Mulzer and G. Rote. Minimum-weight triangulation is NP-hard. \textit{Journal of the ACM}, 55(2): 11:1--11:29 (2008).
	\bibitem{pencils} D. Packer, S. White and A. Williams. A Paper on Pencils: A Pencil and Paper Puzzle - Pencils is NP-Complete. In \textit{Proceedings of the 30th Canadian Conference on Computational Geometry (CCCG)}, pp. 35--41 (2018).
	\bibitem{suguru} L. Robert, D. Miyahara, P. Lafourcade, L. Libralesso and T. Mizuki. Physical zero-knowledge proof and NP-completeness proof of Suguru puzzle. \textit{Information and Computation}, 285(B): 104858 (2022).
	\bibitem{sumplete} S. Ruangwises. Sumplete is Hard, Even with Two Different Numbers. arXiv:2309.07161 (2023). \url{https://arxiv.org/abs/2309.07161}
	\bibitem{toichika} S. Ruangwises. Toichika is NP-Complete. In \textit{Proceedings of the 25th Indonesia-Japan Conference on Discrete and Computational Geometry, Graphs, and Games (IJCDC$G^3$)}, p. 96 (2023).
	\bibitem{3sat} T.J. Schaefer. The complexity of satisfiability problems. In \textit{Proceedings of the 10th Annual ACM Symposium on Theory of Computing (STOC)}, pp. 216--226 (1978).
	\bibitem{ripple} Y. Takenaga, S. Aoyagi, S. Iwata and T. Kasai. Shikaku and Ripple Effect are NP-Complete. \textit{Congressus Numerantium}, 216: 119--127 (2013).
	\bibitem{nonogram} N. Ueda and T. Nagao. NP-completeness Results for NONOGRAM via Parsimonious Reductions. Technical Report TR96-0008, Department of Computer Science, Tokyo Institute of Technology (1996).
	\bibitem{fillmat} A. Uejima and H. Suzuki. Fillmat is NP-Complete and ASP-Complete. \textit{Journal of Information Processing}, 23(3): 310--316 (2015).
	\bibitem{sudoku} T. Yato and T. Seta. Complexity and Completeness of Finding Another Solution and Its Application to Puzzles. \textit{IEICE Trans. Fundamentals}, 86.A(5): 1052--1060 (2003).
\end{thebibliography}

\end{document}